\documentclass[conference]{IEEEtran}
\IEEEoverridecommandlockouts
\usepackage{cite}
\usepackage{amsmath,amssymb,amsfonts}
\usepackage{algorithmic}
\usepackage{graphicx}
\usepackage{textcomp}
\usepackage{longtable}
\usepackage[table]{xcolor}
\usepackage{hyperref}
\def\BibTeX{{\rm B\kern-.05em{\sc i\kern-.025em b}\kern-.08em
    T\kern-.1667em\lower.7ex\hbox{E}\kern-.125em}}
\begin{document}

\title{A Review of Common Online Speaker Diarization Methods\\}

\author{\IEEEauthorblockN{1\textsuperscript{st} Roman Aperdannier}
\IEEEauthorblockA{\textit{Faculty of Business} \\
\textit{University of Applied Science}\\
Ansbach, Germany \\
aperdannier19472@hs-ansbach.de}
\and
\IEEEauthorblockN{2\textsuperscript{nd} Sigurd Schacht}
\IEEEauthorblockA{\textit{Faculty of Business} \\
\textit{University of Applied Science}\\
Ansbach, Germany \\
sigurd.schacht@hs-ansbach.de}
\and
\IEEEauthorblockN{3\textsuperscript{rd} Alexander Piazza}
\IEEEauthorblockA{\textit{Faculty of Business} \\
\textit{University of Applied Science}\\
Ansbach, Germany \\
alexander.piazza@hs-ansbach.de}

}

\maketitle

\begin{abstract}
Speaker diarization provides the answer to the question "who spoke when?" for an audio file. This information can be used to complete audio transcripts for further processing steps. Most speaker diarization systems assume that the audio file is available as a whole. However, there are scenarios in which the speaker labels are needed immediately after the arrival of an audio segment. Speaker diarization with a correspondingly low latency is referred to as online speaker diarization. This paper provides an overview. First the history of online speaker diarization is briefly presented. Next a taxonomy and datasets for training and evaluation are given. In the sections that follow, online diarization methods and systems are discussed in detail. This paper concludes with the presentation of challenges that still need to be solved by future research in the field of online speaker diarization.
\end{abstract}


\section{Introduction}
Speaker diarization is a machine learning task in which the model has the task of assigning audio sequences to the corresponding speakers. Speaker Diarization thus answers the question "who spoke when". In the process of Speaker diarization an audio file is divided into individual audio sequences that are separated by a speaker change or the transition from non-speech to speech. This is important information that is necessary for a fully-fledged transcription of audio files. Speaker diarization in combination with automatic speech recognition (ASR) is therefore used in many transcription scenarios. These scenarios include online meetings, conversations at conferences, earnings reports of public corporations, court proceedings, interviews, social media audios/videos, etc. \cite{park_review_2021}.
In some of these scenarios, it is important that the speaker diarization results are available with a low latency. On the one hand, the transcriber can then make direct adjustments to the transcription. On the other hand, the transcription results can be used directly for further analyses, which can afterwards influence certain actions. For example, whether a company stock should be sold or not based on the statements made in the earnings call \cite{de_castro_mt5b3_2021}. This type of speaker diarization is known as online speaker diarization.

\subsection{Motivation and Background}
There are already two papers \cite{park_review_2021}, \cite{anguera_miro_speaker_2012}  that provide a review of speaker diarization systems. These papers discuss the historical development, evaluation metrics, different diarization methods, common datasets and current use cases of speaker diarization. However, these papers do not explicitly address online speaker diarization methods and systems. This paper aims to fill this gap by conducting a review of common online speaker diarization methods. \\
To give a good overview of online diarization methods, the rest of this paper is structured as follows. The next section will give a general introduction to online diarization. For this purpose, the historical development is presented, the taxonomy for the online diarization systems of this paper is introduced and common evaluation metrics are presented. Next various online diarization systems are briefly summarized. Finally, challenges of online speaker diarization will be presented and an outlook will be given.

\section{Online Speaker Diarization in General}
\begin{figure*}
    \centering
    \includegraphics[width=0.8\textwidth]{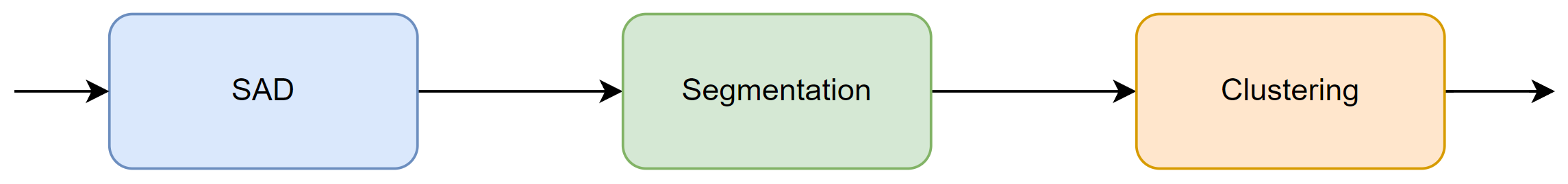}
    \caption{Diarization Pipeline}
    \label{fig:Diarization_Pipeline}
\end{figure*}

Online speaker diarization systems generally work in the same way as offline speaker diarization systems and can be divided into the following sub-tasks:

\begin{itemize}
    \item Speech Activity Detection (SAD): This task enables the system to recognize whether an audio segment contains speech or not.
    \item Segmentation: This task attempts to cut audio segments so that they only contain one speaker.
    \item Clustering: In this step, the audio segments are assigned to the corresponding speakers.
\end{itemize}
This pipeline can also be seen in figure \ref{fig:Diarization_Pipeline}.

However, online diarization systems assume that the input arrives as an audio stream. This means there is not the entire audio file available for speaker diarization. Only the audio segments that have already been annotated can be included in the diarization process of the current audio segment.

\subsection{Historical Development} \label{historical_development}
The first preliminary work on online speaker diarization was published in 1999. In their work, Daben Liu et al. \cite{liu_fast_1999} present an algorithm for recognizing speaker changes in real time. Their system is based on Hidden Markov Models (HMM) in combination with Gaussian Mixture Models (GMM) to define audio classes. A maximum likelihood distance is calculated to recognize a speaker change. \\

A few years later in 2003, Daben Liu et al. \cite{lilt_online_2004} published a paper on online speaker clustering algorithms that have comparable performance to offline speaker clustering algorithms. Two of the newly developed algorithms even outperform the offline hierarchical clustering chosen as a baseline. 

In the following years, the first online speaker diarization systems were introduced. These systems are generally structured as follows, with minor deviations. For SAD, these systems use energy-based SAD systems or consider non-speech as a separate class. For segmentation, the audio file is divided into Mel Frequency Cepstral Coefficients (MFCC) of constant length. GMMs are used as audio and speaker representations. Depending on the system, the GMMs are combined with a Universal Background Model (UBM), to also represent the speaker-independent part of the acoustic features. Agglomerative Hierarchical Clustering (AHC) is the most commonly used clustering algorithm \cite{aronowitz_online_2012}\cite{markov_never-ending_2007}\cite{oku_low-latency_2012}\cite{soldi_adaptive_2015}\cite{vaquero_hybrid_2010}. Some other systems also include the speaker's physical location in the diarization process to further improve the results.
\cite{schmalenstroeer_online_2006}\cite{schmalenstroeer_joint_2007}.

With the introduction of i-vectors \cite{sell_speaker_2014} and d-vectors \cite{variani_deep_2014}, audio and speaker representations in the form of GMMs were replaced in online speaker diarization \cite{dimitriadis_developing_2017}. The use of neural network-based d-vectors in particular led to a leap in performance in speaker diarization \cite{zhang_fully_2019} \cite{fini_supervised_2019}.

End-to-end systems are the latest development in the context of speaker diarization. Here, a single neural network takes over the individual sub tasks of speaker diarization. These end-to-end systems are also used for online speaker diarization. Especially in combination with self-attention, these systems lead to an improvement in performance \cite{coria_overlap-aware_2021}\cite{han_bw-eda-eend_2021}\cite{liang_frame-wise_2023}\cite{wang_end--end_2023} \cite{xue_online_2021}. 

\subsection{Taxonomy of this Work}
In this paper, online diarization systems are divided into two categories. All systems with a modular structure are assigned to the first category. This includes all systems that process at least one sub-task separately. All pure end-to-end systems are assigned to the second category. These include systems that process all sub-tasks with a single machine learning model. Table \ref{table:1} provides an overview of the systems. \textit{Further systems} are not considered in more detail in this paper, but are included in the assigned category.

\begin{table*}[t]
  \centering
    \caption{Overview of online diarization systems}
  \begin{tabular}{| m{7cm} | m{7cm} |}
    \hline
    \rowcolor{lightgray!70} \textbf{Modular Systems} & \textbf{End-to-End Systems} \\
    \hline
    \begin{itemize}
        \item GMM based \cite{markov_never-ending_2007}
        \item I-vector \cite{dimitriadis_developing_2017}
        \item UIS-RNN \cite{zhang_fully_2019} \cite{fini_supervised_2019}
        \item Turn to diarize \cite{xia_turn--diarize_2022}
        \item Further modular systems \cite{aronowitz_online_2012} \cite{oku_low-latency_2012} \cite{schmalenstroeer_online_2006} \cite{schmalenstroeer_joint_2007} \cite{soldi_adaptive_2015} \cite{vaquero_hybrid_2010} \cite{zhang_low-latency_2022}
    \end{itemize} 
    &
    \begin{itemize}
        \item Frame-wise streaming \cite{liang_frame-wise_2023}
        \item Minivox \cite{lin_speaker_2020}
	\item Further EEND systems \cite{han_bw-eda-eend_2021}\cite{wang_end--end_2023} \cite{xue_online_2021}
    \end{itemize}
    \\
    \hline
  \end{tabular}
  \label{table:1}
\end{table*}

\subsection{Metrics}
The metrics described here are not purely online speaker diarization metrics. They are also used in offline speaker diarization.
\subsubsection{DER}
The Diarization Error Rate (DER) \cite{fiscus_rich_2006} is the most common metric to evaluate online diarization systems. The DER is made up of three different errors. These include:
\begin{itemize}
    \item False alarm (FA): When speech is recognized even though there is no speech in the segment
    \item Missed Speech (MS): If speech is not recognized although there is speech in the segment
    \item Speaker Confusion (SC): If the wrong speaker is assigned to a segment.
\end{itemize}
The DER is then calculated from the sum of the errors divided by the duration of the whole audio file as can be seen in Equation (\ref{eq:1}).
\begin{equation}\label{eq:1}
    DER = \frac{FA + MS + SC}{Total Duration of Time}\
\end{equation}
Normally only the DER is specified in the evaluation of diarization systems. But in some cases, the individual error components are also reported.

\subsubsection{JER}
Another metric used in some studies is the Jaccard Error Rate (JER) \cite{ryant_second_2019}. The aim of this metric is to rate each speaker with the same weight. Regardless of the speaker's speaking time. To do this, the error per speaker is first calculated and then divided by speakers speaking time. The error per speaker is the sum of FA and MS as can be seen in Equation (\ref{eq:2}).
\begin{equation}\label{eq:2}
    JER = \frac{1}{N} \sum_{i}^{N_{ref}}\frac{FA_{i}+MS_{i}}{TOTAL_{i}}
\end{equation}

\subsection{Datasets}
The datasets for online speaker diarization are the same as for offline diarization systems. However, the datasets are handed to the online speaker diarization systems as a stream. The following datasets are used most frequently:
\begin{itemize}
    \item CALLHOME: This dataset contains 500 telephone multilingual language sessions with 2 to 7 speakers \cite{canavan_alexandra_callhome_1997}. 
    \item 2003 NIST Rich Transcription: This dataset is a collection of different datasets. It contains a total of 13 hours of annotated multilingual speech with several speakers \cite{fiscus_jonathan_g_2003_2007}. 
    \item DIHARD (I/II/III): These datasets build on each other. The DIHARD datasets are multilingual and contain 1-8 speakers \cite{ryant_first_2018}\cite{ryant_second_2019}\cite{ryant_third_2021}. 
    \item VoxConverse: This dataset contains over 50 hours of multilingual speech extracted from YouTube videos \cite{chung_spot_2020}. 
\end{itemize}

\section{Modular Online Speaker Diarization Systems}
The following sections take a closer look at some of the modular online speaker diarization systems. The systems are arranged in the historical order in which they were published.
\subsection{Online Speaker Diarization with GMMs}
As already described in section \ref{historical_development}, the first online diarization systems were primarily built with GMMs.  GMMs are essentially an iterative soft clustering. First a Gaussian distribution is randomly initialized for each cluster. Next the probabilities of belonging to the Gaussian distribution are then calculated for each data point. Data points that belong to one of the Gaussian distributions with a very high probability (above a certain threshold) are assigned to this Gaussian distribution or cluster. All other data points are assigned to several clusters. After that new Gaussian distributions are calculated from the assigned data points, this process is repeated iteratively until the Gaussian distributions no longer show any major changes. In speaker diarization, MFCCs are taken as data points and a speaker is represented by the mean vectors and the covariance matrices of a GMM. Due to their iterative nature, GMMs can also be used well in online speaker diarization systems \cite{liu_gaussian_2010} \cite{reynolds_robust_1995}. 

As an example of an online diarization system based on GMMs, the work of Markov et. al. \cite{markov_never-ending_2007} will now be described in more detail. Markov et. al. have developed a system that decides for each new speech segment whether it matches a known speaker GMM. If no matching speaker GMM is found, a new speaker GMM is created. Speaker GMMs are deleted if no new speech segment has been assigned to the GMM for a long time. 

The system uses three different GMMs as SAD component. The first represents non-speech. The other GMMs represent voice characteristics of the two biological sexes.

Segmentation is performed using a logic that includes parameters such as minimum segment length (MSL), maximum pause in segment (MPS) and maximum speech in pause (MSP).

In the subsequent clustering component, the system decides whether it is a new speaker or a known speaker. This decision is made using a likelihood ratio. If it is a known speaker, the GMM of this speaker is updated with the new data point. If it is a new speaker, a new speaker GMM is created from the associated gender GMM. This allows the system to integrate new speakers online.
By deleting speaker GMMs that are no longer used, the system is able to process audio endlessly.
From a latency of 3 seconds, the system delivers solid performance on the evaluation dataset with a DER \textless 12\%.

\subsection{Online Diarization with i-vectors}
I-vectors have developed from the problems of GMMs with intersession variability. In this context, intersession variability means that the same speaker sounds different in different recordings. Joint Factor Analysis (JFA) was proposed to counteract this problem \cite{kenny_speaker_2007}. The JFA breaks down the vector of the GMM into individual components. The result is a speaker independent component, speaker dependent component, channel dependent component and a residual component.

In a subsequent study, it was found that the channel dependent component also contains information about the speaker \cite{dehak_front-end_2011}. As a result, the speaker component and the channel component were combined into a common variability matrix. The column weights of this combined matrix are also referred to as i-vectors. In the following years, these i-vectors were used as a representation for audio segments in both online and offline speaker diarization systems.

\begin{figure}[h!]
    \centering
    \includegraphics[width=1\linewidth]{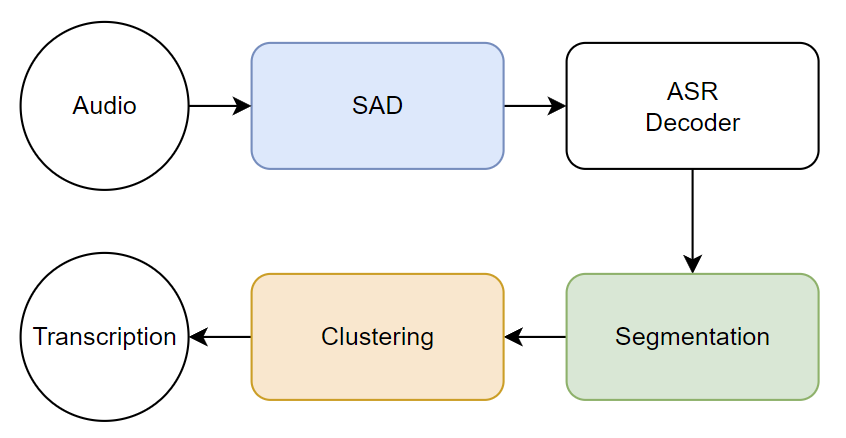}
    \caption{Dimitriadis system architecture}
    \label{fig:i-vec-system}
\end{figure}

Dimitriadis et al. \cite{dimitriadis_developing_2017} have developed an online diarization system that uses i-vectors, among other things. The system consists of a SAD, a segmentation and a clustering component. In addition, an ASR component is integrated to improve the performance of the system. The system architecture of Dimitriadis et. al. is visualized in figure \ref{fig:i-vec-system}. The components of this system will be described in more detail in the following.

The SAD component is based on a neural network and was inspired by the work of Thomas et al. \cite{thomas_improvements_2015}. The output of the SAD component is passed directly to the ASR module. This validates that no words are contained in non-speech segments. Thereby the ASR module can eliminate many false positives from the SAD component.

The segmentation component receives continuous audio segments in the form of cepstral acoustic features and the ASR transcript as input. The audio segment is then split into two segments at each word boundary. Two Gaussians are fitted with the two sub-segments and compared with the BIC algorithm \cite{chen_clustering_1998}. If the two sub-segments differ sufficiently, a speaker change is set here. The additional ASR module ensures that speaker changes are only set at word boundaries.

For the clustering, i-vectors are generated from the homogeneous audio segments. These are then clustered using the x-means algorithm. The x-means is a variation of the k-means algorithm \cite{pelleg_x-means_2000}. The x-means has a linear complexity proportional to the number of audio segments to be clustered. However, in order to ensure online diarization, long audio files cannot be clustered on the entire history. Therefore, Dimitriadis et al. introduce an active window to limit the history. However, this means that clustering of two different active windows is no longer consistent. To solve this problem, the system provides a fast speaker label based on the clustering of the active window. A more accurate speaker label is then delivered with a slightly higher latency. The accurate label is generated using a reconciliation algorithm \cite{church_speaker_2017}, which minimizes the hamming distance of the speaker labels between two adjacent active windows. This approach gives the end user the option of receiving speaker labels with a short latency and updating them with the more accurate labels at a later point in time.

\subsection{Supervised Online Clustering - UIS RNN}
As already described in Section \ref{historical_development} Historical development, the current speaker diarization systems are based on neural networks (NN). There is one approach of replacing the entire system with a single neuronal network. These end-to-end systems are discussed in more detail in the next section \ref{EEND systems}. Another approach is to replace individual sub-tasks with trainable neural networks. 
In the online diarization system by Zhang et. al. \cite{zhang_fully_2019}, the focus is on the use of a fully supervised clustering component. This system is described in more detail below.

As SAD, two simple Gaussian distributions are used to filter out audio segments that do not contain speech. Subsequently, overlapping audio segments are cut. For the segments, representations are generated in the form of d-vectors \cite{variani_deep_2014}. These d-vectors are used as input for supervised clustering. 
The algorithm unbounded interleaved-state recurrent neural network (UIS RNN) was developed as a clustering component for this system. UIS RNN receives three sets as input in the training scenario:

\begin{itemize}
    \item \(X = (x_{1}, x_{2}, x_{3}, \text{…}, x_{T})\) where each \(x_{t}\) is the d-vector of an audio segment
    \item \(Y = (y_{1}, y_{2}, y_{3}, \text{…}, y_{T})\) where each \(y_{t}\) is the speaker-id of the corresponding audio segment
    \item \(Z = (z_{2}, z_{3}, \text{…}, z_{T})\) where \(z_{t}=1\) if there is a speaker change at time \(t\). In all other cases, \(z_{t}=0\)
\end{itemize}

The aim of UIS RNN is to represent the combined probability of \(p(X,Y,Z)\). UIS RNN can therefore be split into three individual components.

\subsubsection{Speaker Change \(p(z_{t}|z_{[t-1]})\)}
The speaker change component indicates how likely it is that there will be a speaker change at time \(t\). In the implementation of UIS RNN, this component is implemented as a coin flip for simplification. 

\subsubsection{Speaker Assignment \(p(y_{t} | z_{t}, y_{[t-1]})\)}
This component models the probability of which speaker is assigned after a speaker change. UIS RNN uses the Chinese restaurant process (CRP) \cite{blei_distance_2011} for this task. This ensures that a speaker who has already spoken often is more likely to be assigned than a speaker who has spoken less frequently. The probability of a new speaker entering the conversation is represented by a constant probability. The reason for this is that in a single domain, the probability of a new speaker joining the conversation is fairly constant. For example, a new speaker is very unlikely to join a phone call, but very likely to join a movie.

\subsubsection{Sequence Generation \(p(x_{t}| x_{[t-1]}, y_{[t]})\)}
The RNN variant Gated Recurrent Unit (GRU) \cite{cho_learning_2014} is used to implement this component. The aim of this component is to model the probability of the embedding \(x_{t}\) based on the embedding \(x_{[t-1]}\) and the speaker label \(y_{t}\). To do this, UIS RNN creates separate GRU instances for each speaker. Each instance has a state \(h_{t}\) at time \(t\) depending on the speaker \(y_{t}\). The output of the entire RNN is \(m_{t}\). The current sequence \(x_{t}\) is inferred from \(m_{t}\).

The system is trained by maximizing the logarithm of the combined probability \(P(X,Y,Z)\). The GRU hidden states of the RNN and the number of times a speaker has already spoken in the CRP are updated in the training process by a greedy MAP algorithm after each new audio sequence. As a result, UIS RNN works online and can generate a speaker label with a short latency as soon as a new audio segment arrives. 
With the described supervised approach of UIS-RNN, the algorithm achieves better results than state-of-the-art spectral offline clustering algorithms in their evaluation. Although UIS-RNN works online.

\subsection{Supervised Online Clustering - Turn to Diarize}
Supervised speaker diarization systems need a lot of training data in order to deliver good results. The annotation of audio data is time consuming and therefore cost intensive. As a rule, you can expect to spend 2 hours on a 10 minute audio file \cite{xia_turn--diarize_2022}. In their work, Xia et. al. \cite{xia_turn--diarize_2022} present a system that can be trained on the basis of speaker turn labels. For this purpose, $<st>$  tokens are inserted into the ASR transcript of the audio file at each speaker turn. This means that an exact timestamp no longer needs to be set during annotation. This speeds up labeling many times over. In addition, the semantic information in the audio data can be processed better.

The system from Xia et al. is structured as follows. A transformer transducer model takes over the ASR and speaker turn detection. The detected segments are then fed into a Long short-term memory model (LSTM) in order to calculate the corresponding d-vectors. An online variant of spectral clustering is used to cluster the d-vectors, which also includes the speaker turns for decision making. The individual system components are described below.

\subsubsection{Speaker Turn Detection}
A recurrent neural network transducer (RNN-T) is used for this purpose. This is a supervised ASR model consisting of an audio encoder, a label encoder and an neural network for generating the final output sequences. To increase the speed of the model, the Transformer Transducer (T-T) variant of the RNN-T and a bigram label encoder are used in their work. During training, the model receives transcripts with speaker turn tokens $<st>$ as target output and log-mel audio features as input. At inference time, the entire output of the model is ignored, except for the $<st>$ tokens with their associated timestamp. In addition, a confidence score is calculated for the $<st>$ token, which is important for clustering.

\subsubsection{Speaker Encoder}
An LSTM is used as the speaker encoder, which provides a d-vector as output. The model functions independently of the ASR script. As an input the model receives audio segments, which are separated by speaker turns. The model only uses 75\% of the audio segments and ignores the information at the segment boundaries. This reduces the risk of errors being carried over from the speaker turn detection component.

\subsubsection{Spectral Clustering}
The spectral clustering algorithm with an additional speaker turn constraint is used as the clustering component. A constraint matrix is calculated for this \(Q \epsilon R^{NxN}\), where \(N\) is the number of audio segments. The matrix is filled as follows:
\begin{itemize}
    \item If there is a speaker turn between the adjacent segments \(i\) and \(i+1\)  with a confidence score greater than a threshold, these segments are labeled as \textit{cannot-link (-1)}.
    \item If there is no speaker turn between adjacent segments, these are labeled as \textit{must-link (+1)} in the constraint matrix.
    \item Non-adjacent segments are assigned \(Q_{ij} = 0\).
\end{itemize}
For values in the constraint matrix Q \textgreater{} 0, the similarity between the segments is increased in the clustering process. For values \textless{} 0, the similarity is reduced. 

RNN-T and LSTM are streaming models. The bottleneck of the system is therefore the clustering component. However, the number of segments can be significantly reduced by using turn wise segments. This leads to a leap in clustering performance. As a result, spectral clustering can be performed after each arrival of a new segment and the entire system can be used in online speaker diarization scenarios.

\section{End-to-End Online Speaker Diarization} \label{EEND systems}
As mentioned in the previous section, supervised online diarization systems also include so called end-to-end systems. These train a single neural network to solve all sub tasks of speaker diarization. Two of these end-to-end systems are described in more detail below.

\subsection{Frame-wise Streaming End-to-End Speaker Diarization} \label{FSEEND}
\begin{figure}[ht]
    \centering
    \includegraphics[width=0.7\linewidth]{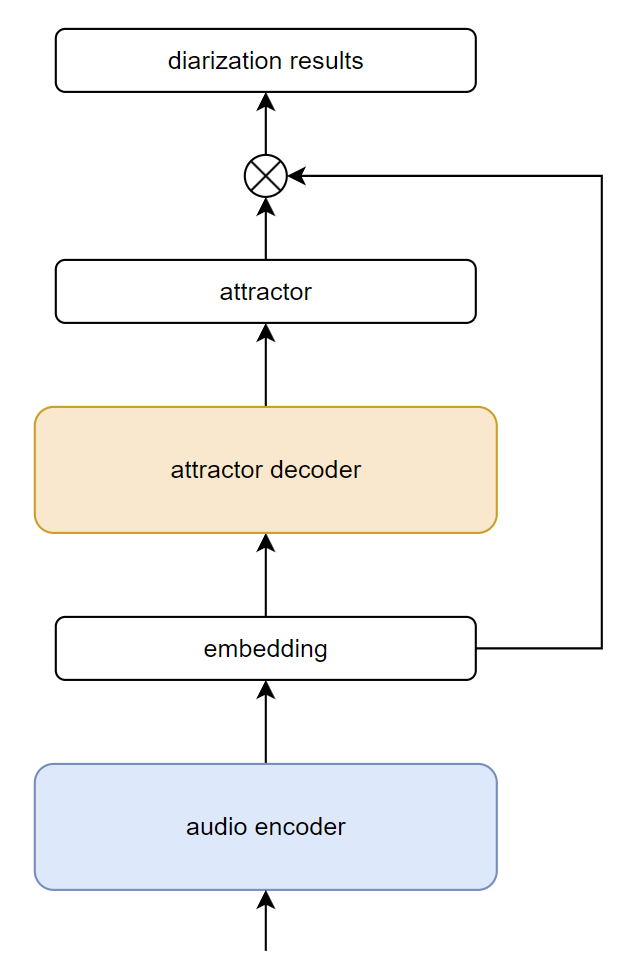}
    \caption{System architecture of FS-EEND}
    \label{fig:fseend-system}
\end{figure}

In their work, Liang et al. \cite{liang_frame-wise_2023} present an online speaker diarization system that processes the audio stream frame by frame and delivers the corresponding diarization results in real time. The system consists of an audio encoder and an attractor decoder. In this case, an attractor is the representation of a speaker. Finally, diarization results are generated by a similarity comparison of the attractors with the audio embeddings. The system architecture can be seen in figure \ref{fig:fseend-system}.
The decoder is implemented with non-autoregressive self-attention. On the one hand, self-attention ensures that the speaker labels remain consistent over time. On the other hand, self-attention makes it possible to better distinguish the speakers. In order to achieve better performance, a look-ahead mechanism is added to the system. This mechanism receives some future frames as input and thus makes it possible to recognize new unknown speakers in real time. The architecture of the whole system is analyzed in more detail below.

\subsubsection{Embedding Encoder}
The embedding encoder consists of a linear layer and a masked transformer encoder. The frame-wise input sequence \(X=(x_{1}, ..., x_{1}, ..., x_{T})\) is binary masked so that the self-attention module in the transformer cannot include any future information.
\subsubsection{Look-ahead Mechanism}
In most online speaker diarization scenarios, a low latency is acceptable. In addition, the quality of the embeddings and attractors can be significantly increased with a few future frames. For these reasons, it was decided to implement a look-ahead mechanism in this system. This is realized in the form of a one dimensional convolution along the time axis \cite{liang_frame-wise_2023}. The latency (number of future frames) is controlled by the kernel size of the convolution.
\subsubsection{Attractor Decoder}
The decoder is based on self-attention. Self-attention based systems have a longer memory than, for example, an LSTM-based system. This means that known speakers can be correctly assigned over a longer period of time. The difference to tasks such as sequence generation is that the results in the form of attractors do not have to be output in a specific order. For this reason, a non-autoregressive attractor decoder is used in this work, which can output several attractors in parallel.

In general, EEND systems do not cope well if the audio files for inference contain more speakers than in training data \cite{horiguchi_online_2023}. For this reason, the maximum number of speakers is limited to 4 in the work of Liang et al. \cite{liang_frame-wise_2023}. 

The decoder receives the embedding of the current frame as input and decides which attractors are updated with the embedding. The input must always be different for each attractor, therefore a positional encoding is added to the embedding. 
To obtain an attractor for the current frame \(a_{s,t}\), the decoder needs two sources of information. Firstly, the previous attractors of the speaker \(a_{s,t'}\). This is achieved by the masked frame self-attention (MFSA) module. Secondly, the attractors of the other speakers to the current frame \(a_{s',t}\) in order to sharpen the distance to the other attractors. This task is performed by the decoder's cross-attractor self-attention (CASA) module. The two self-attention modules are combined in a feed forward network, which provides the attractors as output.
Finally, the speaker labels \(y\) can be calculated using the inner product of the attractors \(A_{t} = [a_{1,t}, ... , a_{s,t}, ... , a_{S,t}]\) and the embedding \(e_{t}\).
For training, the sum of two loss functions is minimized. The first loss function is the binary cross entropy (BCE) of predicted speaker labels \(\hat{Y}\) and the ground truth \(Y\).

\begin{equation} \label{eq:3}
    Loss_{d}=\frac{1}{T}\sum_{t=1}^{T}BCE(\hat{y_{t}},y_{t})
\end{equation}

The second loss function is the embedding similarity loss, which is calculated by the mean squared error of the cosine similarity between two embeddings and the cosine similarity of the corresponding speaker labels.

\begin{equation} \label{eq:4}
    Loss_{e}=\frac{1}{TxT}\sum_{j=1}^{T}\sum_{k=1}^{T}MSE(\left \langle e_{j},e_{k} \right \rangle,\left \langle y_{j}, y_{k} \right \rangle)
\end{equation}
The latency of the system can be controlled by the kernel size of the look-ahead component. FS-EEND achieves better results than the selected comparison systems from a latency time of 1s upwards.

\subsection{Online Learning Minivox}
In their work, Lin et al. \cite{lin_speaker_2020} present a benchmark for speaker diarization systems based on online learning. In addition they develop an online learning speaker diarization system. In the following, both the benchmark and the online diarization system will be examined in more detail.

The Minivox Benchmark converts classic speaker diarization datasets into audio streams. To do this, random speaker \textit{n} and associated audio sequences \textit{m} are attached to each other. The parameter \textit{p} can be used to control how many of the speaker labels the system receives as feedback. The parameter \textit{p} thus simulates a real user who does not always provide feedback. A result can look like this, for example:
\begin{itemize}
    \item \(X = [1,1,1,2,2,2,1,1,3,3,3,3, …] \)
    \item \(Y = [1,\_,\_,2,\_,\_,\_,1,\_,\_3,\_,…]\ for\ p\ =\ 0.3 \)
\end{itemize}
Training can be carried out either with or without an oracle. The oracle specifies the maximum number of speakers as initial input. The performance of the system is measured with the DER.

The presented Online Speaker Diarization System is based on the concept of the Contextual Bandit \cite{agrawal_thompson_2013}. For each decision option, the bandit has an arm that stands for an unknown reward. The bandit tries to find a good trade-off between exploiting known options and discovering new, possibly better options. In case of the contextual bandit, the bandit is given a context (e.g. audio embeddings) that he can incorporate into his decision. The context in this work is the maximum number of speakers. 

If the online diarization system starts with the maximum number of speakers \textit{N}, the bandit initially receives \textit{N} arms, as well as an additional arm for \textit{no speech}. Without this information, the bandit starts with one arm for \textit{new speaker} and one arm for \textit{no speech}. The system then begins the learning and inference process. The following cases must be distinguished:

\begin{itemize}
    \item The system selects \textit{new speaker} and the user confirms the selection: The system initializes a \textit{new speaker}.
    \item The system selects \textit{no speech} and the user objects to the selection: The system initializes a \textit{new speaker}.
    \item The system selects speaker \(n_{x}\) and the user objects with \textit{new speaker}: The system initializes a new speaker by copying speaker \(n_{x}\).
\end{itemize}

In addition, it may be the case that the user does not provide any feedback. A self-supervision module has been added to the system for this purpose. This clusters the audio segments that have already been labeled and the current audio segment. Then the cluster label is used as a substitute for the user feedback. Online variants of GMM, k-means and KNN are used as cluster algorithms. The authors are aware that these perform poorly compared to there offline variants, but want to ensure the online suitability of the system.

The presented online diarization system has a comparatively poor performance. However, the online learning scenario poses a particular challenge, as the system never gets to see the real labels of the audio segments unless the user confirms an assigned speaker label.

\section{Challenges of Online Speaker Diarization}
Online speaker diarization has to face a number of challenges. These include the trade-off between accuracy and latency, missing training data and problems with multispeaker in case of the EEND systems. The various challenges are examined in more detail below.

\subsection{Tradeoff between Accuracy and Latency}
In many current online diarization systems, it can be observed that the error is reduced if a higher latency is accepted. The reason for this is that more audio information can be included in the decision if a larger input sequence is available. Depending on the system, this trade-off is handled differently. Coria et al. \cite{coria_overlap-aware_2021} use an active window that is controlled by the parameter \(\lambda\). The larger \(\lambda\) is selected, the higher the latency and the lower the DER. Morrone et al. \cite{morrone_low-latency_2022} use a similar approach in their work with a CSS window based on the sliding window of Chen et al. \cite{chen_continuous_2020}. Here it can be also observed that the DER decreases when the window is enlarged. Liang et al. \cite{liang_frame-wise_2023} show the system some future frames via the look-ahead mechanism. As described in section \ref{FSEEND}, the latency is controlled via the kernel size of the convolution that makes the future frames available. Dimitriadis et al. \cite{dimitriadis_developing_2017} solve the problem by returning a fast label and later delivering a more accurate label. Further examples of the accuracy latency tradeoff can be found in \cite{han_bw-eda-eend_2021} \cite{wang_speaker_2022} \cite{xue_online_2021}.

However, it can be observed that the DER no longer drops significantly above a certain latency. This sweet spot latency differs from system to system. For Coria et al. \cite{coria_overlap-aware_2021} it is 3s, for Morrone et al. \cite{morrone_low-latency_2022} it is a window size of 15s. Nevertheless, there is a need for further research to decouple latency from accuracy.

\subsection{Training Data}
Current supervised online and offline diarization systems need a lot of high quality training data to deliver good results. Such training data is only available to a limited extent in English and is usually associated with high costs \cite{canavan_alexandra_callhome_1997} \cite{fiscus_jonathan_g_2003_2007} \cite{cieri_christopher_fisher_2005}. For some languages, there is still no corresponding training data for speaker diarization. An approach such as Xia et al. \textit{Turn to Diarize} \cite{xia_turn--diarize_2022} attempts to simplify the annotation of the training data. Lin et al.  \cite{lin_speaker_2020} try to perform the training process through online learning parallel to inference time. These are good approaches, but further research and work is needed to reduce this problem.

\subsection{Multispeaker in EEND Systems}
Many end-to-end online diarization systems can only process a limited number of speakers. Horiguchi et al. \cite{horiguchi_online_2023} state in their work that end-to-end online diarization systems always have problems when the number of speakers for inference is higher than the number of speakers in the training process. In their work, they also propose a solution by adding blockwise clustering to the end-to-end system. However, they move away from the actual core idea of end-to-end systems, which is to solve the entire speaker diarization with a single neural network. Also, the maximum number of speakers in online speaker diarization cannot be determined in advance, as the audio file arrives as a stream and additional speakers may be added at a later point in time.  This is a research gap that needs to be closed in the future in order to be able to handle a flexible number of speakers for inference.

\section{Conclusion}
Online speaker diarization is a research topic that has been dealt with for a long time. The first systems were created with GMMs. With the introduction of i-vectors, GMM-based systems were largely replaced. A short time later, embeddings of audio segments were developed in the form of d-vectors, which are generated by neural networks. However, not only audio representations were replaced by trainable modules. Other components of the speaker diarization pipeline, such as the clustering, have also been replaced by supervised approaches. The latest innovations are end-to-end online diarization systems. These take the approach of replacing the entire diarization pipeline with a single trainable model.

All these developments have continuously improved online speaker diarization. However, online speaker diarization is a challenging topic. The limited amount of input data makes it difficult to undercut the error rate of comparable offline diarization systems. Thus, a tradeoff between accuracy and latency must always be made. Rare training data is a problem for both offline and online speaker diarization. In end-to-end systems, the maximum number of speakers is implicitly limited by the training data used. Solutions still need to be developed for this too. 

In summary, this paper provides a good overview of the topic of online speaker diarization. Also this paper shows that online speaker diarization is a current and flourishing topic that still offers a lot of potential for further research.

\bibliographystyle{IEEEtran}
\bibliography{references}

\begin{thebibliography}{10}
\providecommand{\url}[1]{#1}
\csname url@samestyle\endcsname
\providecommand{\newblock}{\relax}
\providecommand{\bibinfo}[2]{#2}
\providecommand{\BIBentrySTDinterwordspacing}{\spaceskip=0pt\relax}
\providecommand{\BIBentryALTinterwordstretchfactor}{4}
\providecommand{\BIBentryALTinterwordspacing}{\spaceskip=\fontdimen2\font plus
\BIBentryALTinterwordstretchfactor\fontdimen3\font minus \fontdimen4\font\relax}
\providecommand{\BIBforeignlanguage}[2]{{%
\expandafter\ifx\csname l@#1\endcsname\relax
\typeout{** WARNING: IEEEtran.bst: No hyphenation pattern has been}%
\typeout{** loaded for the language `#1'. Using the pattern for}%
\typeout{** the default language instead.}%
\else
\language=\csname l@#1\endcsname
\fi
#2}}
\providecommand{\BIBdecl}{\relax}
\BIBdecl

\bibitem{park_review_2021}
\BIBentryALTinterwordspacing
T.~J. Park, N.~Kanda, D.~Dimitriadis, K.~J. Han, S.~Watanabe, and S.~Narayanan, ``A {Review} of {Speaker} {Diarization}: {Recent} {Advances} with {Deep} {Learning},'' Nov. 2021, arXiv:2101.09624 [cs, eess]. [Online]. Available: \url{http://arxiv.org/abs/2101.09624}
\BIBentrySTDinterwordspacing

\bibitem{de_castro_mt5b3_2021}
P.~A.~L. de~Castro, ``mt5b3: {A} {Framework} for {Building} {Autonomous} {Traders},'' \emph{CoRR}, 2021.

\bibitem{anguera_miro_speaker_2012}
\BIBentryALTinterwordspacing
X.~Anguera~Miro, S.~Bozonnet, N.~Evans, C.~Fredouille, G.~Friedland, and O.~Vinyals, ``\BIBforeignlanguage{en}{Speaker {Diarization}: {A} {Review} of {Recent} {Research}},'' \emph{\BIBforeignlanguage{en}{IEEE Transactions on Audio, Speech, and Language Processing}}, vol.~20, no.~2, pp. 356--370, Feb. 2012. [Online]. Available: \url{http://ieeexplore.ieee.org/document/6135543/}
\BIBentrySTDinterwordspacing

\bibitem{liu_fast_1999}
D.~Liu and F.~Kubala, ``Fast speaker change detection for broadcast news transcription and indexing,'' in \emph{Sixth {European} conference on speech communication and technology}.\hskip 1em plus 0.5em minus 0.4em\relax Citeseer, 1999.

\bibitem{lilt_online_2004}
D.~Lilt and F.~Kubala, ``Online speaker clustering,'' in \emph{2004 {IEEE} {International} {Conference} on {Acoustics}, {Speech}, and {Signal} {Processing}}, vol.~1, May 2004, pp. I--333, iSSN: 1520-6149.

\bibitem{aronowitz_online_2012}
H.~Aronowitz, Y.~A. Solewicz, and O.~Toledo-Ronen, ``Online two speaker diarization,'' in \emph{Odyssey 2012-{The} {Speaker} and {Language} {Recognition} {Workshop}}, 2012.

\bibitem{markov_never-ending_2007}
\BIBentryALTinterwordspacing
K.~Markov and {Satoshi Nakamura}, ``\BIBforeignlanguage{en}{Never-ending learning system for on-line speaker diarization},'' in \emph{\BIBforeignlanguage{en}{2007 {IEEE} {Workshop} on {Automatic} {Speech} {Recognition} \& {Understanding} ({ASRU})}}.\hskip 1em plus 0.5em minus 0.4em\relax The Westin Miyako Kyoto: IEEE, 2007, pp. 699--704. [Online]. Available: \url{http://ieeexplore.ieee.org/document/4430197/}
\BIBentrySTDinterwordspacing

\bibitem{oku_low-latency_2012}
T.~Oku, S.~Sato, A.~Kobayashi, S.~Homma, and T.~Imai, ``Low-latency speaker diarization based on {Bayesian} information criterion with multiple phoneme classes,'' in \emph{2012 {IEEE} {International} {Conference} on {Acoustics}, {Speech} and {Signal} {Processing} ({ICASSP})}.\hskip 1em plus 0.5em minus 0.4em\relax IEEE, 2012, pp. 4189--4192.

\bibitem{soldi_adaptive_2015}
\BIBentryALTinterwordspacing
G.~Soldi, C.~Beaugeant, and N.~Evans, ``\BIBforeignlanguage{en}{Adaptive and online speaker diarization for meeting data},'' in \emph{\BIBforeignlanguage{en}{2015 23rd {European} {Signal} {Processing} {Conference} ({EUSIPCO})}}.\hskip 1em plus 0.5em minus 0.4em\relax Nice: IEEE, Aug. 2015, pp. 2112--2116. [Online]. Available: \url{http://ieeexplore.ieee.org/document/7362757/}
\BIBentrySTDinterwordspacing

\bibitem{vaquero_hybrid_2010}
C.~Vaquero, O.~Vinyals, and G.~Friedland, ``A hybrid approach to online speaker diarization.'' in \emph{{InterSpeech}}, 2010, pp. 2638--2641.

\bibitem{schmalenstroeer_online_2006}
\BIBentryALTinterwordspacing
J.~Schmalenstroeer and R.~Haeb-Umbach, ``\BIBforeignlanguage{en}{Online speaker change detection by combining {BIC} with microphone array beamforming},'' in \emph{\BIBforeignlanguage{en}{Interspeech 2006}}.\hskip 1em plus 0.5em minus 0.4em\relax ISCA, Sep. 2006, pp. paper 1078--Wed1FoP.2--0. [Online]. Available: \url{https://www.isca-speech.org/archive/interspeech\_2006/schmalenstroeer06\_interspeech.html}
\BIBentrySTDinterwordspacing

\bibitem{schmalenstroeer_joint_2007}
------, ``Joint speaker segmentation, localization and identification for streaming audio,'' in \emph{Eighth {Annual} {Conference} of the {International} {Speech} {Communication} {Association}}.\hskip 1em plus 0.5em minus 0.4em\relax Citeseer, 2007.

\bibitem{sell_speaker_2014}
\BIBentryALTinterwordspacing
G.~Sell and D.~Garcia-Romero, ``\BIBforeignlanguage{en}{Speaker diarization with plda i-vector scoring and unsupervised calibration},'' in \emph{\BIBforeignlanguage{en}{2014 {IEEE} {Spoken} {Language} {Technology} {Workshop} ({SLT})}}.\hskip 1em plus 0.5em minus 0.4em\relax South Lake Tahoe, NV, USA: IEEE, Dec. 2014, pp. 413--417. [Online]. Available: \url{http://ieeexplore.ieee.org/document/7078610/}
\BIBentrySTDinterwordspacing

\bibitem{variani_deep_2014}
\BIBentryALTinterwordspacing
E.~Variani, X.~Lei, E.~McDermott, I.~L. Moreno, and J.~Gonzalez-Dominguez, ``\BIBforeignlanguage{en}{Deep neural networks for small footprint text-dependent speaker verification},'' in \emph{\BIBforeignlanguage{en}{2014 {IEEE} {International} {Conference} on {Acoustics}, {Speech} and {Signal} {Processing} ({ICASSP})}}.\hskip 1em plus 0.5em minus 0.4em\relax Florence, Italy: IEEE, May 2014, pp. 4052--4056. [Online]. Available: \url{http://ieeexplore.ieee.org/document/6854363/}
\BIBentrySTDinterwordspacing

\bibitem{dimitriadis_developing_2017}
D.~Dimitriadis and P.~Fousek, ``Developing {On}-{Line} {Speaker} {Diarization} {System}.'' in \emph{Interspeech}, 2017, pp. 2739--2743.

\bibitem{zhang_fully_2019}
\BIBentryALTinterwordspacing
A.~Zhang, Q.~Wang, Z.~Zhu, J.~Paisley, and C.~Wang, ``Fully {Supervised} {Speaker} {Diarization},'' Feb. 2019, arXiv:1810.04719 [cs, eess, stat]. [Online]. Available: \url{http://arxiv.org/abs/1810.04719}
\BIBentrySTDinterwordspacing

\bibitem{fini_supervised_2019}
\BIBentryALTinterwordspacing
E.~Fini and A.~Brutti, ``Supervised online diarization with sample mean loss for multi-domain data,'' Nov. 2019, arXiv:1911.01266 [cs, eess] version: 3. [Online]. Available: \url{http://arxiv.org/abs/1911.01266}
\BIBentrySTDinterwordspacing

\bibitem{coria_overlap-aware_2021}
\BIBentryALTinterwordspacing
J.~M. Coria, H.~Bredin, S.~Ghannay, and S.~Rosset, ``Overlap-aware low-latency online speaker diarization based on end-to-end local segmentation,'' Sep. 2021, arXiv:2109.06483 [cs, eess]. [Online]. Available: \url{http://arxiv.org/abs/2109.06483}
\BIBentrySTDinterwordspacing

\bibitem{han_bw-eda-eend_2021}
\BIBentryALTinterwordspacing
E.~Han, C.~Lee, and A.~Stolcke, ``{BW}-{EDA}-{EEND}: {Streaming} {End}-to-{End} {Neural} {Speaker} {Diarization} for a {Variable} {Number} of {Speakers},'' in \emph{{ICASSP} 2021 - 2021 {IEEE} {International} {Conference} on {Acoustics}, {Speech} and {Signal} {Processing} ({ICASSP})}, Jun. 2021, pp. 7193--7197, arXiv:2011.02678 [cs, eess]. [Online]. Available: \url{http://arxiv.org/abs/2011.02678}
\BIBentrySTDinterwordspacing

\bibitem{liang_frame-wise_2023}
\BIBentryALTinterwordspacing
D.~Liang, N.~Shao, and X.~Li, ``Frame-wise streaming end-to-end speaker diarization with non-autoregressive self-attention-based attractors,'' Sep. 2023, arXiv:2309.13916 [cs, eess] version: 1. [Online]. Available: \url{http://arxiv.org/abs/2309.13916}
\BIBentrySTDinterwordspacing

\bibitem{wang_end--end_2023}
\BIBentryALTinterwordspacing
W.~Wang and M.~Li, ``End-to-end {Online} {Speaker} {Diarization} with {Target} {Speaker} {Tracking},'' Oct. 2023, arXiv:2310.08696 [cs, eess]. [Online]. Available: \url{http://arxiv.org/abs/2310.08696}
\BIBentrySTDinterwordspacing

\bibitem{xue_online_2021}
\BIBentryALTinterwordspacing
Y.~Xue, S.~Horiguchi, Y.~Fujita, S.~Watanabe, and K.~Nagamatsu, ``Online {End}-to-{End} {Neural} {Diarization} with {Speaker}-{Tracing} {Buffer},'' Mar. 2021, arXiv:2006.02616 [cs, eess]. [Online]. Available: \url{http://arxiv.org/abs/2006.02616}
\BIBentrySTDinterwordspacing

\bibitem{xia_turn--diarize_2022}
\BIBentryALTinterwordspacing
W.~Xia, H.~Lu, Q.~Wang, A.~Tripathi, Y.~Huang, I.~L. Moreno, and H.~Sak, ``Turn-to-{Diarize}: {Online} {Speaker} {Diarization} {Constrained} by {Transformer} {Transducer} {Speaker} {Turn} {Detection},'' Jan. 2022, arXiv:2109.11641 [cs, eess] version: 3. [Online]. Available: \url{http://arxiv.org/abs/2109.11641}
\BIBentrySTDinterwordspacing

\bibitem{zhang_low-latency_2022}
\BIBentryALTinterwordspacing
Y.~Zhang, Q.~Lin, W.~Wang, L.~Yang, X.~Wang, J.~Wang, and M.~Li, ``Low-{Latency} {Online} {Speaker} {Diarization} with {Graph}-{Based} {Label} {Generation},'' Jun. 2022, arXiv:2111.13803 [cs, eess]. [Online]. Available: \url{http://arxiv.org/abs/2111.13803}
\BIBentrySTDinterwordspacing

\bibitem{lin_speaker_2020}
\BIBentryALTinterwordspacing
B.~Lin and X.~Zhang, ``Speaker {Diarization} as a {Fully} {Online} {Learning} {Problem} in {MiniVox},'' Oct. 2020, arXiv:2006.04376 [cs, stat] version: 3. [Online]. Available: \url{http://arxiv.org/abs/2006.04376}
\BIBentrySTDinterwordspacing

\bibitem{fiscus_rich_2006}
J.~G. Fiscus, J.~Ajot, M.~Michel, and J.~S. Garofolo, ``The rich transcription 2006 spring meeting recognition evaluation,'' in \emph{Machine {Learning} for {Multimodal} {Interaction}: {Third} {International} {Workshop}, {MLMI} 2006, {Bethesda}, {MD}, {USA}, {May} 1-4, 2006, {Revised} {Selected} {Papers} 3}.\hskip 1em plus 0.5em minus 0.4em\relax Springer, 2006, pp. 309--322.

\bibitem{ryant_second_2019}
\BIBentryALTinterwordspacing
N.~Ryant, K.~Church, C.~Cieri, A.~Cristia, J.~Du, S.~Ganapathy, and M.~Liberman, ``The {Second} {DIHARD} {Diarization} {Challenge}: {Dataset}, task, and baselines,'' Jun. 2019, arXiv:1906.07839 [cs, eess]. [Online]. Available: \url{http://arxiv.org/abs/1906.07839}
\BIBentrySTDinterwordspacing

\bibitem{canavan_alexandra_callhome_1997}
\BIBentryALTinterwordspacing
{Canavan, Alexandra}, {Graff, David}, and {Zipperlen, George}, ``{CALLHOME} {American} {English} {Speech},'' 1997, artwork Size: 1830160 KB Pages: 1830160 KB. [Online]. Available: \url{https://catalog.ldc.upenn.edu/LDC97S42}
\BIBentrySTDinterwordspacing

\bibitem{fiscus_jonathan_g_2003_2007}
\BIBentryALTinterwordspacing
{Fiscus, Jonathan G.}, {Doddington, George R.}, {Le, Audrey}, {Sanders, Greg}, {Przybocki, Mark}, and {Pallett, David}, ``2003 {NIST} {Rich} {Transcription} {Evaluation} {Data},'' Aug. 2007, artwork Size: 2097152 KB Pages: 2097152 KB. [Online]. Available: \url{https://catalog.ldc.upenn.edu/LDC2007S10}
\BIBentrySTDinterwordspacing

\bibitem{ryant_first_2018}
N.~Ryant, K.~Church, C.~Cieri, A.~Cristia, J.~Du, S.~Ganapathy, and M.~Liberman, ``First {DIHARD} challenge evaluation plan,'' \emph{tech. Rep.}, 2018, publisher: Linguistic Data Consortium.

\bibitem{ryant_third_2021}
\BIBentryALTinterwordspacing
N.~Ryant, P.~Singh, V.~Krishnamohan, R.~Varma, K.~Church, C.~Cieri, J.~Du, S.~Ganapathy, and M.~Liberman, ``The {Third} {DIHARD} {Diarization} {Challenge},'' Apr. 2021, arXiv:2012.01477 [cs, eess]. [Online]. Available: \url{http://arxiv.org/abs/2012.01477}
\BIBentrySTDinterwordspacing

\bibitem{chung_spot_2020}
\BIBentryALTinterwordspacing
J.~S. Chung, J.~Huh, A.~Nagrani, T.~Afouras, and A.~Zisserman, ``Spot the conversation: speaker diarisation in the wild,'' in \emph{Interspeech 2020}, Oct. 2020, pp. 299--303, arXiv:2007.01216 [cs, eess]. [Online]. Available: \url{http://arxiv.org/abs/2007.01216}
\BIBentrySTDinterwordspacing

\bibitem{liu_gaussian_2010}
\BIBentryALTinterwordspacing
J.~Liu, D.~Cai, and X.~He, ``\BIBforeignlanguage{en}{Gaussian {Mixture} {Model} with {Local} {Consistency}},'' \emph{\BIBforeignlanguage{en}{Proceedings of the AAAI Conference on Artificial Intelligence}}, vol.~24, no.~1, pp. 512--517, Jul. 2010, number: 1. [Online]. Available: \url{https://ojs.aaai.org/index.php/AAAI/article/view/7659}
\BIBentrySTDinterwordspacing

\bibitem{reynolds_robust_1995}
\BIBentryALTinterwordspacing
D.~Reynolds and R.~Rose, ``Robust text-independent speaker identification using {Gaussian} mixture speaker models,'' \emph{IEEE Transactions on Speech and Audio Processing}, vol.~3, no.~1, pp. 72--83, Jan. 1995, conference Name: IEEE Transactions on Speech and Audio Processing. [Online]. Available: \url{https://ieeexplore.ieee.org/abstract/document/365379}
\BIBentrySTDinterwordspacing

\bibitem{kenny_speaker_2007}
\BIBentryALTinterwordspacing
P.~Kenny, G.~Boulianne, P.~Ouellet, and P.~Dumouchel, ``\BIBforeignlanguage{en}{Speaker and {Session} {Variability} in {GMM}-{Based} {Speaker} {Verification}},'' \emph{\BIBforeignlanguage{en}{IEEE Transactions on Audio, Speech and Language Processing}}, vol.~15, no.~4, pp. 1448--1460, May 2007. [Online]. Available: \url{http://ieeexplore.ieee.org/document/4156203/}
\BIBentrySTDinterwordspacing

\bibitem{dehak_front-end_2011}
N.~Dehak, P.~Kenny, R.~Dehak, P.~Dumouchel, and P.~Ouellet, ``Front-{End} {Factor} {Analysis} for {Speaker} {Verification},'' \emph{Audio, Speech, and Language Processing, IEEE Transactions on}, vol.~19, pp. 788--798, Jun. 2011.

\bibitem{thomas_improvements_2015}
\BIBentryALTinterwordspacing
S.~Thomas, G.~Saon, M.~V. Segbroeck, and S.~S. Narayanan, ``\BIBforeignlanguage{en}{Improvements to the {IBM} speech activity detection system for the {DARPA} {RATS} program},'' in \emph{\BIBforeignlanguage{en}{2015 {IEEE} {International} {Conference} on {Acoustics}, {Speech} and {Signal} {Processing} ({ICASSP})}}.\hskip 1em plus 0.5em minus 0.4em\relax South Brisbane, Queensland, Australia: IEEE, Apr. 2015, pp. 4500--4504. [Online]. Available: \url{http://ieeexplore.ieee.org/document/7178822/}
\BIBentrySTDinterwordspacing

\bibitem{chen_clustering_1998}
S.~S. Chen and P.~S. Gopalakrishnan, ``Clustering via the {Bayesian} information criterion with applications in speech recognition,'' in \emph{Proceedings of the 1998 {IEEE} {International} {Conference} on {Acoustics}, {Speech} and {Signal} {Processing}, {ICASSP}'98 ({Cat}. {No}. {98CH36181})}, vol.~2.\hskip 1em plus 0.5em minus 0.4em\relax IEEE, 1998, pp. 645--648.

\bibitem{pelleg_x-means_2000}
D.~Pelleg, A.~W. Moore, and {others}, ``X-means: {Extending} k-means with efficient estimation of the number of clusters.'' in \emph{Icml}, vol.~1, 2000, pp. 727--734.

\bibitem{church_speaker_2017}
\BIBentryALTinterwordspacing
K.~Church, W.~Zhu, J.~Vopicka, J.~Pelecanos, D.~Dimitriadis, and P.~Fousek, ``Speaker diarization: {A} perspective on challenges and opportunities from theory to practice,'' in \emph{2017 {IEEE} {International} {Conference} on {Acoustics}, {Speech} and {Signal} {Processing} ({ICASSP})}, Mar. 2017, pp. 4950--4954, iSSN: 2379-190X. [Online]. Available: \url{https://ieeexplore.ieee.org/abstract/document/7953098}
\BIBentrySTDinterwordspacing

\bibitem{blei_distance_2011}
\BIBentryALTinterwordspacing
D.~M. Blei and P.~I. Frazier, ``\BIBforeignlanguage{en}{Distance {Dependent} {Chinese} {Restaurant} {Processes}},'' Aug. 2011, arXiv:0910.1022 [stat]. [Online]. Available: \url{http://arxiv.org/abs/0910.1022}
\BIBentrySTDinterwordspacing

\bibitem{cho_learning_2014}
\BIBentryALTinterwordspacing
K.~Cho, B.~van Merrienboer, C.~Gulcehre, D.~Bahdanau, F.~Bougares, H.~Schwenk, and Y.~Bengio, ``\BIBforeignlanguage{en}{Learning {Phrase} {Representations} using {RNN} {Encoder}-{Decoder} for {Statistical} {Machine} {Translation}},'' Sep. 2014, arXiv:1406.1078 [cs, stat]. [Online]. Available: \url{http://arxiv.org/abs/1406.1078}
\BIBentrySTDinterwordspacing

\bibitem{horiguchi_online_2023}
\BIBentryALTinterwordspacing
S.~Horiguchi, S.~Watanabe, P.~Garcia, Y.~Takashima, and Y.~Kawaguchi, ``Online {Neural} {Diarization} of {Unlimited} {Numbers} of {Speakers} {Using} {Global} and {Local} {Attractors},'' \emph{IEEE/ACM Transactions on Audio, Speech, and Language Processing}, vol.~31, pp. 706--720, 2023, arXiv:2206.02432 [cs, eess]. [Online]. Available: \url{http://arxiv.org/abs/2206.02432}
\BIBentrySTDinterwordspacing

\bibitem{agrawal_thompson_2013}
S.~Agrawal and N.~Goyal, ``Thompson sampling for contextual bandits with linear payoffs,'' in \emph{International conference on machine learning}.\hskip 1em plus 0.5em minus 0.4em\relax PMLR, 2013, pp. 127--135.

\bibitem{morrone_low-latency_2022}
\BIBentryALTinterwordspacing
G.~Morrone, S.~Cornell, D.~Raj, L.~Serafini, E.~Zovato, A.~Brutti, and S.~Squartini, ``Low-{Latency} {Speech} {Separation} {Guided} {Diarization} for {Telephone} {Conversations},'' Oct. 2022, arXiv:2204.02306 [eess] version: 2. [Online]. Available: \url{http://arxiv.org/abs/2204.02306}
\BIBentrySTDinterwordspacing

\bibitem{chen_continuous_2020}
\BIBentryALTinterwordspacing
Z.~Chen, T.~Yoshioka, L.~Lu, T.~Zhou, Z.~Meng, Y.~Luo, J.~Wu, X.~Xiao, and J.~Li, ``Continuous speech separation: dataset and analysis,'' May 2020, arXiv:2001.11482 [cs, eess]. [Online]. Available: \url{http://arxiv.org/abs/2001.11482}
\BIBentrySTDinterwordspacing

\bibitem{wang_speaker_2022}
\BIBentryALTinterwordspacing
Q.~Wang, C.~Downey, L.~Wan, P.~A. Mansfield, and I.~L. Moreno, ``Speaker {Diarization} with {LSTM},'' Jan. 2022, arXiv:1710.10468 [cs, eess, stat]. [Online]. Available: \url{http://arxiv.org/abs/1710.10468}
\BIBentrySTDinterwordspacing

\bibitem{cieri_christopher_fisher_2005}
\BIBentryALTinterwordspacing
{Cieri, Christopher}, {Graff, David}, {Kimball, Owen}, {Miller, Dave}, and {Walker, Kevin}, ``Fisher {English} {Training} {Part} 2, {Speech},'' Apr. 2005, artwork Size: 29643984 KB Pages: 29643984 KB. [Online]. Available: \url{https://catalog.ldc.upenn.edu/LDC2005S13}
\BIBentrySTDinterwordspacing

\end{thebibliography}

\end{document}